# REPORT ON THE PCAPAC 2000 WORKSHOP

R. Bacher[*], Deutsches Elektronen-Synchrotron DESY, D-22603 Hamburg, Germany

## Abstract

In October 2000, the third PCaPAC (PCs and Particle Accelerator Controls) workshop took place at DESY. This paper presents a summary of the workshop. The workshop reviewed existing and new PC-based accelerator control systems, from small-scale to large-scale installations. It demonstrated convincingly the advantage of modern, commercial mass-market products used for accelerator controls. Disadvantages of these technologies were reported as well. Large-scale PC systems inherently bring administrative concerns into the picture. In this vein, special emphasis was given to system administration for distributed systems. A major topic of the workshop was the integration of different control system approaches as well as the integration of different platforms within the same control system. In particular, PC-based concepts offer the simple opportunity to interface to commercial SCADA systems. In addition, large emphasis was given to the presentation of future developments including the next network trends to data exchange via SOAP and XML.

## 1 INTRODUCTION

The PCaPAC (PCs and Particle Accelerator Controls) workshop began in October 1996 at DESY and was designed to be a biennial workshop. The second workshop was held in January 1999 at KEK and the third took place in October 2000 again at DESY[†]. PCaPAC is scheduled to be held in the "non-ICALEPCS" years.

PCaPAC is not in conflict to ICALEPCS, although it has a significant overlap in topics with ICALEPCS due to the common focus on accelerator control systems. However, PCaPAC focus particularly onto the impact of PCs to particle accelerator controls. The usage of PCs bring especial advantages as well as disadvantages. Therefore, the workshop provides a platform to exchange ideas and experience in the dedicated field of PC-related technologies where trends are changing rapidly and where the pace at which hardware as well as software evolve is very fast.

PCaPAC is designed to be a workshop. Following this approach, no parallel sessions were scheduled to give everybody a chance to attend. The morning sessions were dedicated to oral presentation categorized according to the announced workshop topics. The afternoon sessions were split; a poster session was followed by an open-discussion session guided by a chairman. These discussion sessions picked up open questions left from the morning sessions and provided a forum to discuss special topics raised by the conference attendees or the chairman.

Participation in PCaPAC 2000 reached an all-time high of 127 registered attendees from 43 different institutes and 17 countries.

## 2 WORKSHOP TOPICS

### 2.1 Scientific Programme

The scientific programme of the workshop was categorized according the following topics:

Controlling Accelerator Subsystems and Experiments (37 contributions)
- running systems: including commercial or SCADA systems
- peripheral systems: including archiving, alarm, sequencing etc.
- process control: including OPC
- data acquisition
- real-time solutions

Interfacing Accelerator Hardware (2 contributions)
- IO control
- field busses
- device drivers etc.

Accelerator Control Objects and Components (14 contributions)
- object-oriented design and implementation involving OLE, DCOM, ActiveX, Java, Java Beans, CORBA, C++ etc.

Integrating Different Systems (6 contributions)
- operating systems and platforms
- commercial systems, SCADA systems, in-house systems, web-based systems

Control System Architecture (15 contributions)
- methods of distributing control and data exchange mechanisms
- layering models for data flow

System Administration and Project Management (7 contributions)
- impact of system administration on machine operation
- control system project management using PC-based tools

Operator Interface (3 contributions)
- man-machine interface
- ergonomics etc.

Future Trends and Technologies (8 contributions)
- XML, 64-bit processors, WAP, data warehousing and mining, blue tooth
- accommodating trends in old systems

In total, 92 contributions (oral: 31, poster: 58, tutorial: 3) were presented (see Fig.1).

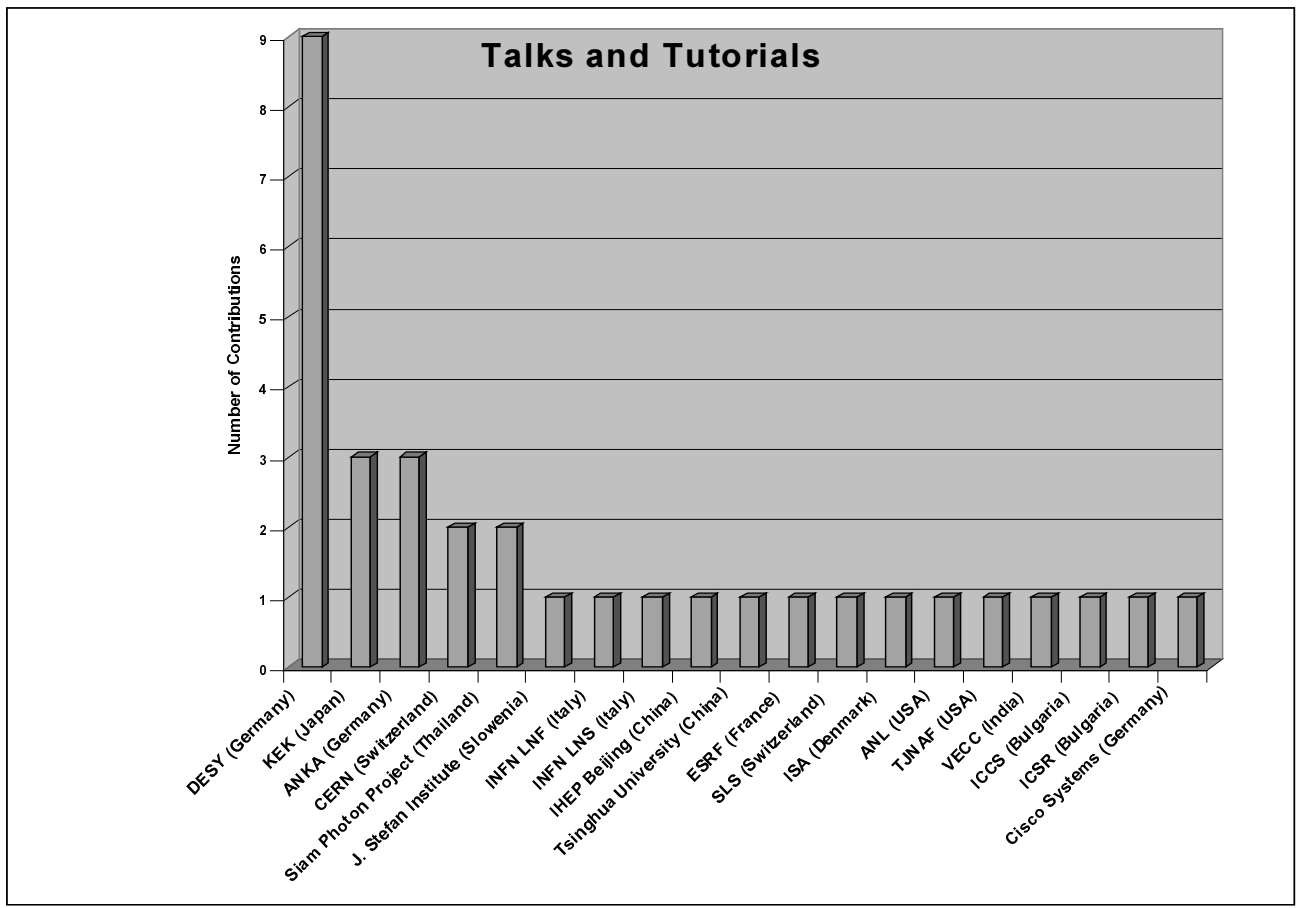

Figure 1: Number of talks and tutorials versus presenting institutes

*2.2 Status and Progress Reports*

At PCaPAC 2000, many running accelerators of which the control-system infrastructure was based entirely or in part on PCs were presented. Among these were small systems built and maintained by a few people (e.g. the storage rings ASTRID and ELISA of the ISA Storage Ring Facilities at the University of Aarhus in Denmark), medium-scale systems (e.g. the ANKA synchrotron light source in Karlsruhe in Germany, and accelerators at KEK in Japan) and probably the largest PC-based control systems for the HERA, PETRA and DORIS storage rings and their injectors at DESY in Hamburg (Germany). The volume and success of development work was manifest in a series of corresponding status reports.

The idea of the control system for the synchrotron radiation light source ANKA is to provide a user-friendly control system where "user" means a physicist operating the accelerator and not a computer expert [1]. The control system contains around 500 physical devices (power supplies, vacuum pumps, beam position monitors, RF generators, etc.) The device I/O is handled by self-sufficient micro-controller boards, which connect to a standard LonWorks field bus network. Each branch of the network is attached to a PC. Device servers running on these PCs map the devices and their properties onto 1500 objects and make them remotely available using CORBA. On the client side, the objects are wrapped into specially developed Java Beans called Abeans [2], which provide also a rich framework of tools that clients always need. This allows applications to be written easily, even by nonprogrammers.

COACK (Component Oriented Accelerator Control Kernel) [3] being developed at KEK in Japan in joint venture with IT-industry is a general-purpose kernel for accelerator controls based on Microsoft's COM/DCOM foundation. It consists of different components like a class builder, data cache, message exchanger (peer-to-peer and publish-subscribe type communication), accelerator virtual machine, operation support (scheduling and polling), database (SQL, XML, Excel) and data/command logging, own registry and diagnostic tools (session manager and status display). In addition, support to integrate complete SCADA systems and lab automation software like LabView or HPVEE is provided.

## 2.3 System Administration Concerns

The distributed nature of PC-based control systems is manifest in the important role that is played by system administration.

The automated installation for Linux PCs (reconstruction farms, work group server and desktop workstations) is done at DESY with YaST, the administration tool of the SuSE Linux distribution. In spring 2000, SuSE has introduced the successor tool YaST2 [4]. DESY took the chance to influence the YaST2 development in an early stage and have SuSE integrate features to turn YaST2 into a network based configuration management. YaST2 takes a fully modular approach to configuration, installation, and administration of the system. It comes with its own scripting language ycp (YaST control protocol), which has an interface to call other code, like C, perl, or shell script. YaST2 supports remote, network transparent configuration, installation, and administration of a PC or a group of PCs by users or group administrators. The configuration is stored on a network database. Existing configurations can be cloned to set up clusters of identical PCs. A hierarchical administration structure can be realised.

To manage the Windows-based HERA console workstations at DESY, NetInstall is used for system and application management [5]. The idea is to use the same tool for the HERA consoles as for the desktop PCs at DESY. The benefit is threefold, (1) sharing of knowledge, (2) exporting control software to all office PCs at DESY and (3) importing standard software from central application support. About 50 dedicated control PCs are located in various control rooms. The update of DLLs and other important software is done at logon time. In addition, a check of the control application at launch time for newer versions is performed.

## 2.4 Integrating Different Control Systems

It is a common and unavoidable experience that different control systems have to be interfaced. Complete commercial SCADA systems as well as in-house developed control systems have to be integrated into other accelerator control systems.

An example is the EPICS-to-TINE translator [6] at DESY. TINE is the dominant accelerator control system protocol at HERA [7]. The translator server resides directly on the EPICS IOC. The EPICS names are mapped to TINE device names. The Channel Access Protocol is by-passed and the process variables are accessed directly using the database access layer including the necessary data type conversions.

## 2.5 Future Trends and Technologies

PCaPAC has been seen to be worthwhile in discussions involving the technologies of the fast evolving internet.

A. Pace from CERN [8] demonstrated impressively the usage of the Web as a platform-independent application platform and the potential of SOAP (Simple Object Access Protocol) as a Web-wide application protocol based on HTTP and XML. He recommended the integration of the control systems into the Web to make Web technologies to be the core and not the border of the control systems. Furthermore, he pointed out the possibilities which rendering software provides. This will open control systems not only to traditional Web browsers but also mobile phones, WebTV or any internet enabled device.

## 3 ACKNOWLEDGEMENT

The author would like to thank P. Duval, I. Nikodem, J. Maaß, R. Schmidt, R. Schröder and W. Schütte for their great enthusiasm and effort organizing the PCaPAC 2000 workshop.

---

\* for the PCaPAC 2000 organizing committee
† http://desyntwww.desy.de/pcapac/